\documentclass[twocolumn,superscriptaddress,pra,reprint,longbibliography]{revtex4-1}
\usepackage{graphicx}
\usepackage{amssymb}
\usepackage{amsmath}
\usepackage{epsfig}
\usepackage{color}
\usepackage{natbib}
\usepackage{mathtools}
\usepackage[colorlinks,linkcolor=blue,anchorcolor=blue,citecolor=blue,urlcolor=blue]{hyperref}

\begin{document}

\title{Indistinguishability Induced Classical-to-Nonclassical Transition of
Photon Statistics}
\author{Ao Shen}
\author{Bo Du}
\author{Ze-Hao Wang}
\author{Yang Dong}
\author{Xiang-Dong Chen}
\author{Guang-Can Guo}
\author{Fang-Wen Sun}
\email{fwsun@ustc.edu.cn}
\affiliation{Key Lab of Quantum Information, Chinese Academy of Sciences, School of
physics, University of Science and Technology of China, Hefei, 230026, P.R.
China}
\affiliation{Synergetic Innovation Center of Quantum Information $\&$ Quantum Physics,
University of Science and Technology of China, Hefei, 230026, P.R. China}
\date{\today}

\begin{abstract}
Photon statistics is one of the key properties of photon state for the study of quantum
coherence and quantum information techniques. Here, we discussed the photon
indistinguishability induced bunching effect which can significantly change
photon statistics. Through the measurement of the second-order degree of
coherence of a mixed photon state composed by a single-photon state and a
weak coherent state, the statistical transition from a classical to nonclassical behavior was experimentally demonstrated by modifying the
indistinguishability of the two photon states. The study will help to
understand and control the photon statistics with a new manner. It also
indicates that the photon indistinguishability is a key parameter for multi-partite quantum coherence.
\end{abstract}

\maketitle

The photon statistics is a fundamental property of quantum optical field,
which has been the basic of quantum coherence \cite{QO} and recently
developed optical quantum information techniques \cite%
{RevModPhys.79.135,OBrien1567,10.1038/nphoton.2011.35,RevModPhys.74.145}. It
also has been well applied in quantum super-resolution microscopy
\cite{Cui13,Monti} to achieve nanoscale resolution. Generally, the photon
statistics is mainly determined by the number of the emitters and the
process of the photon-matter interaction. For example, a single photon \cite%
{0034-4885-68-5-R04,Kimble77,10.1038/nature01086,10.1038/nnano.2012.262} can
be generated from a single quantum emitter, which is a key photon
source for quantum communication \cite%
{PhysRevLett.67.661,PhysRevLett.68.557,PhysRevLett.108.130503} and quantum
computation \cite{RevModPhys.79.135,OBrien1567}. Multi-photon state from
nonlinear optical process has been applied to demonstrate quantum
entanglement, quantum computation and high sensitivity quantum metrology
\cite{10.1038/nphoton.2011.35,Nagata07,SUNEPL,Xiang11}. In experiment, the
statistics of a photon state can be modified by post-selection measurement
\cite{FaselNJP}, interaction with atoms \cite{10.1038/17295,Birnbaum05,Fan05}
and interference with another photon state \cite{HOM,SA,Compos89,Sun07,Liu09}%
. In the interference process, besides the phase modulation, the
indistinguishability of photons is also a key parameter. In principle, the
indistinguishability of photons will induce photon bunching and stimulated
emission \cite{Sun07,Liu09}. It has been the basic of multi-photon interference \cite%
{PhysRevA.67.022301,PhysRevLett.96.240502} for scalable optical quantum
information techniques, lasers and stimulated emission depletion microscopy
\cite{Hell94}. \

Experimentally, the photon statistics can be evaluated with the
Hanbury-Brown-Twiss (HBT) interferometer \cite{HBT} to get the second-order
degree of coherence \cite{QO}, $g^{(2)}(0)$. The values of $g^{(2)}(0)$
demonstrate different photon statistical behaviors. A coherent light source
\cite{QO} with a Poissonian distribution of photon numbers has a ${%
g^{(2)}}\left( 0\right) $ of $1$. For a classical optical field, ${g^{(2)}}%
\left( 0\right) \geq 1$. For example, a thermal state shows ${g^{(2)}}\left(
0\right) =2$, demonstrating a photon bunching behavior. While, with a photon anti-bunching behavior, ${g^{(2)}}\left( 0\right) <1$, it is a typical
quantum optical field, such as a perfect single-photon source with ${g^{(2)}}%
\left( 0\right) =0$. For a nonclassical $N$-photon number state, ${g^{(2)}}%
\left( 0\right) =(N-1)/N<1$. However, it is much more complicated for a
photon state composing of different photon number states where the photon
indistinguishability induced bunching factor will modify the photon
statistics. For an $N$-photon state, when they are indistinguishable, the
photon bunching effect will show an $N!$ coefficient because of the permutation symmetry of a boson system \cite{Sun07PRA}. While, for partial
indistinguishable cases, the photon bunching coefficient will drop. For total
distinguishable cases, there is no bunching effect. Therefore the
indistinguishability\ induced bunching factor will modify the amplitude of
each $N$-photon state and significantly change the photon statistical behavior.
\begin{figure}[b]
\centering
\includegraphics[width=7.5cm]{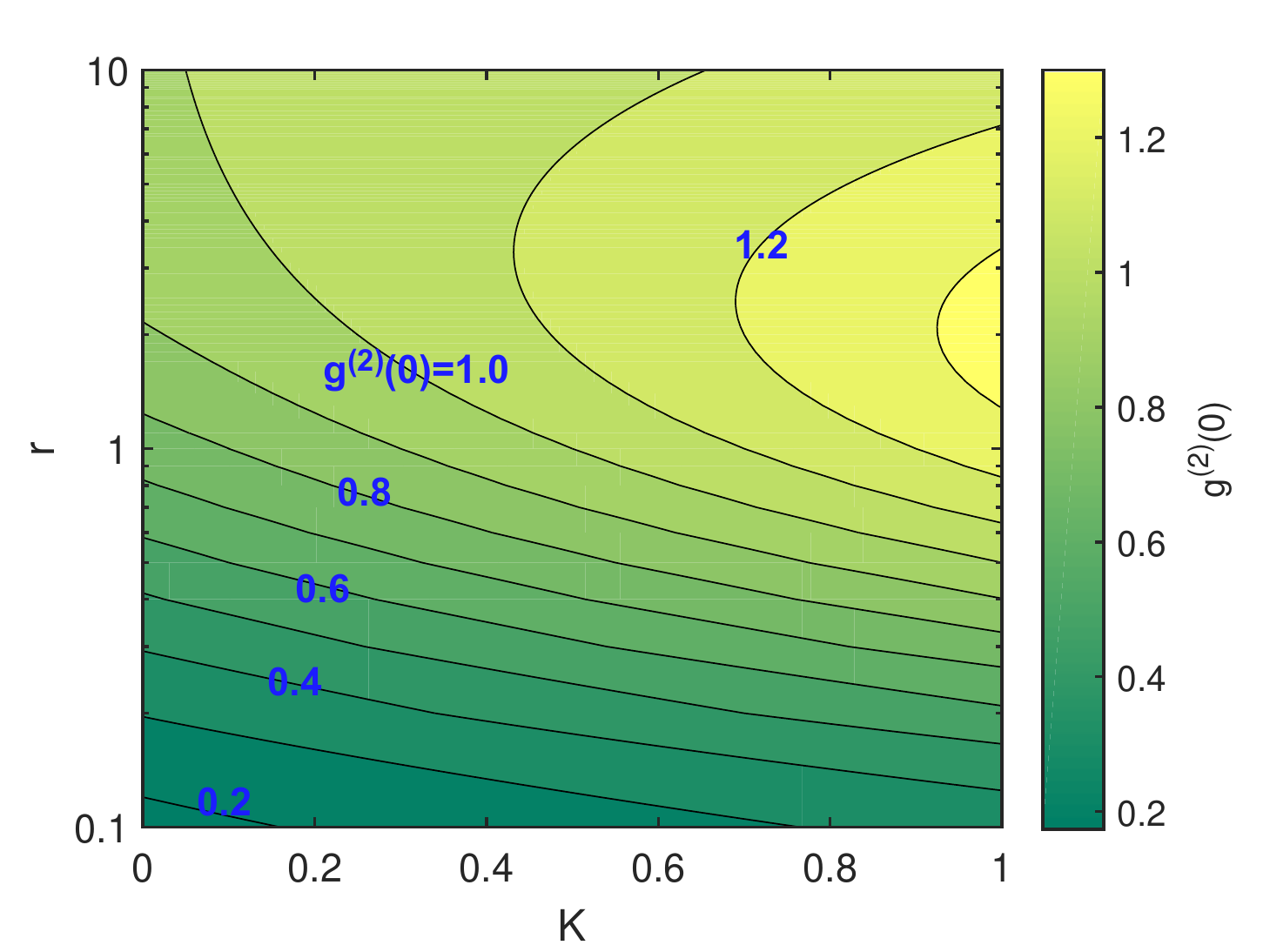}
\caption{$g^{(2)}(0)$ of the mixed photon state versus mixing ratio $r$ and photon
indistinguishability $K$.}
\label{Fig1}
\end{figure}
In this work, we studied the photon statistics by changing the
indistinguishability of photons based on the measurement of ${g^{(2)}}\left(
0\right) $. With a photon interference process \cite{HOM,SA,Sun07}, we
experimentally demonstrated that the photon statistics can be changed from
the bunching behavior ($g^{(2)}(0)>1$) to anti-bunching behavior ($%
g^{(2)}(0)<1$) by modifying the indistinguishability of photons from $0.86$ to $%
0$, realizing the transition from a classical to nonclassical optical field. This
study will help to understand and control the photon statistics with a new
manner for quantum optical coherence and quantum information applications.

In the study of photon indistinguishability and statistics, we consider the
interference of a single-photon state and a weak coherent state.
Theoretically, the single photon state should be $\left\vert 1\right\rangle $%
. However, in a practical case with imperfect photon coupling and detection,
the single-photon state with ${g^{(2)}}\left( 0\right) =0$ can be written as
${\rho _{s}}=\left( {1-\eta }\right) \left\vert 0\right\rangle \langle
0|+\eta \left\vert 1\right\rangle \langle 1|$, where $\left\vert
0\right\rangle $ is the vacuum state and $\eta $ is the mean photon number.
And the weak coherent state with mean photon number of ${\left\vert \alpha
\right\vert ^{2}}$($\ll 1$) is $\left\vert \alpha \right\rangle $ with ${%
g^{(2)}}\left( 0\right) =1$. Then the single photon state and the weak
coherent state is mixed with a mixing ratio of $r={{\left\vert \alpha
\right\vert {}^{2}/}\eta }$. The indistinguishability ($K$) \cite%
{PhysRevA.79.013824} corresponds to the overlapping of the two photon
states. When the photons from these two sources are totally distinguishable ($K=0$%
), the mixed photon state is:
\begin{eqnarray}
\rho _{k=0} &=&{\rho _{s}}\otimes \left\vert \alpha \right\rangle \langle
\alpha |  \notag \\
&=&\left( {1-\eta }\right) \left\vert 0\right\rangle \langle 0|\otimes
\left\vert \alpha \right\rangle \langle \alpha |+\eta \left\vert
1\right\rangle \langle 1|\otimes \left\vert \alpha \right\rangle \langle
\alpha |{\text{.}}  \label{rho1}
\end{eqnarray}%
The ${g^{(2)}}\left( 0\right) $ of the mixed photon state is:
\begin{equation}
{g^{(2)}}\left( 0\right) =\frac{{{{\left\vert \alpha \right\vert }^{2}}%
\left( {2\eta +{{\left\vert \alpha \right\vert }^{2}}}\right) }}{{{{\left( {%
\eta +{{\left\vert \alpha \right\vert }^{2}}}\right) }^{2}}}}=\frac{{{r^{2}}%
+2r}}{{{{\left( {1+r}\right) }^{2}}}}{\text{.}}  \label{gtwo1}
\end{equation}%
For a classical mixing ($K=0$) of the two photon states with ${g^{(2)}}\left(
0\right) =1$ and ${g^{(2)}}\left( 0\right) =0$, we can always find that ${%
g^{(2)}}\left( 0\right) <1$, demonstrating a photon anti-bunching behavior.
However, when the photons from these two sources are totally overlapping and
indistinguishable ($K=1$), the photon indistinguishability will induce quite
different result. When $K=1$, the state of the mixed photons can be written
as:
\begin{equation}
\rho _{k=1}={\rho _{s}}\otimes \left\vert \alpha \right\rangle \langle
\alpha |=C\left[ {\left( {1-\eta }\right) \left\vert \alpha \right\rangle
\langle \alpha |+\eta \left\vert {\alpha ^{\prime }}\right\rangle \langle
\alpha ^{\prime }|}\right] {\text{,}}  \label{rho2}
\end{equation}%
where $\left\vert {\alpha ^{\prime }}\right\rangle ={a^{\dag }}\left\vert
\alpha \right\rangle $ and $C$ is a normalization number. Here, the
single-photon added coherent state ${a^{\dag }}\left\vert \alpha
\right\rangle $ \cite{Zavatta} shows different amplitude with $\left\vert
1\right\rangle \langle 1|\otimes \left\vert \alpha \right\rangle \langle
\alpha |$ in Eq. (\ref{rho1}) because of the photon indistinguishability
induced photon bunching factor. Similarly, the value of ${g^{(2)}}\left(
0\right) $ is%
\begin{equation}
{g^{(2)}}\left( 0\right) =\frac{{\left( {{{\left\vert \alpha \right\vert }%
^{4}}+4\eta {{\left\vert \alpha \right\vert }^{2}}+4\eta {{\left\vert \alpha
\right\vert }^{4}}+\eta {{\left\vert \alpha \right\vert }^{6}}}\right)
\left( {1+\eta {{\left\vert \alpha \right\vert }^{2}}}\right) }}{{{{\left( {%
\eta +{{\left\vert \alpha \right\vert }^{2}}+2\eta {{\left\vert \alpha
\right\vert }^{2}}+\eta {{\left\vert \alpha \right\vert }^{4}}}\right) }^{2}}%
}}\text{.}  \label{gtwo2}
\end{equation}%
When the mean photon numbers of both sources are much smaller than $1$ ($%
\eta ,{\left\vert \alpha \right\vert ^{2}}\ll 1$), Eq.~(\ref{gtwo2}) can be
simplified as:
\begin{equation}
{g^{\left( 2\right) }}\left( 0\right) =\frac{{{{\left\vert \alpha
\right\vert }^{4}}+4\eta {{\left\vert \alpha \right\vert }^{2}}}}{{{{\left( {%
\eta +{{\left\vert \alpha \right\vert }^{2}}}\right) }^{2}}}}{\text{ }}={%
\text{ }}\frac{{{r^{2}}+4r}}{{{{\left( {1+r}\right) }^{2}}}}\text{.}
\end{equation}%
It is easy to find that ${g^{(2)}}\left( 0\right) >1$ when $r>0.5$. In these
cases, the photon indistinguishability induced bunching effect significantly
changes the photon statistics.

For partially indistinguishable cases with $0<K<1$, the mixed photon state
can be
\begin{eqnarray}
{\rho _{K}} &\propto &{\text{ }}\left( {1-\eta }\right) \left\vert \alpha
\right\rangle \langle \alpha |+\frac{\eta }{{1+{{\left\vert \alpha
\right\vert }^{2}}}}\left\vert 1\right\rangle \langle 1|  \notag \\
&&+\frac{{{{\left\vert \alpha \right\vert }^{2}}\eta (1+K)}}{{1+{{\left\vert
\alpha \right\vert }^{2}}}}\left\vert {1,1^{\prime }}\right\rangle \langle
1,1^{\prime }|...\text{,}  \label{rho3}
\end{eqnarray}%
where the coherent state is represented by number states with the higher
order terms dropped since ${\left\vert \alpha \right\vert ^{2}}\ll 1$. $%
\left\vert {1,1^{\prime }}\right\rangle $ represents the state of two
photons with partial indistinguishability with an amplitude enhancement of $%
K $ over the distinguishable case. Also, the ${g^{(2)}}\left( 0\right) $ of
the mixed photon state can be deducted as
\begin{equation}
{g^{2}}(0)=\frac{{{r^{2}}+2(1+K)r}}{{{{\left( {1+r}\right) }^{2}}}}\text{.}
\label{gtwo3}
\end{equation}%
Here the photon statistics highly depends on the value of
indistinguishability. Fig.\ref{Fig1} shows the ${g^{(2)}}\left( 0\right) $
of the mixed photon state with different $r$ and $K$. The transition from
the photon antibunching (${g^{(2)}}\left( 0\right) <1$) to bunching (${%
g^{(2)}}\left( 0\right) >1$) behavior can be realized by increasing $K$ with
some $r$.

\begin{figure}[t]
\centering
\includegraphics[width=7.5cm]{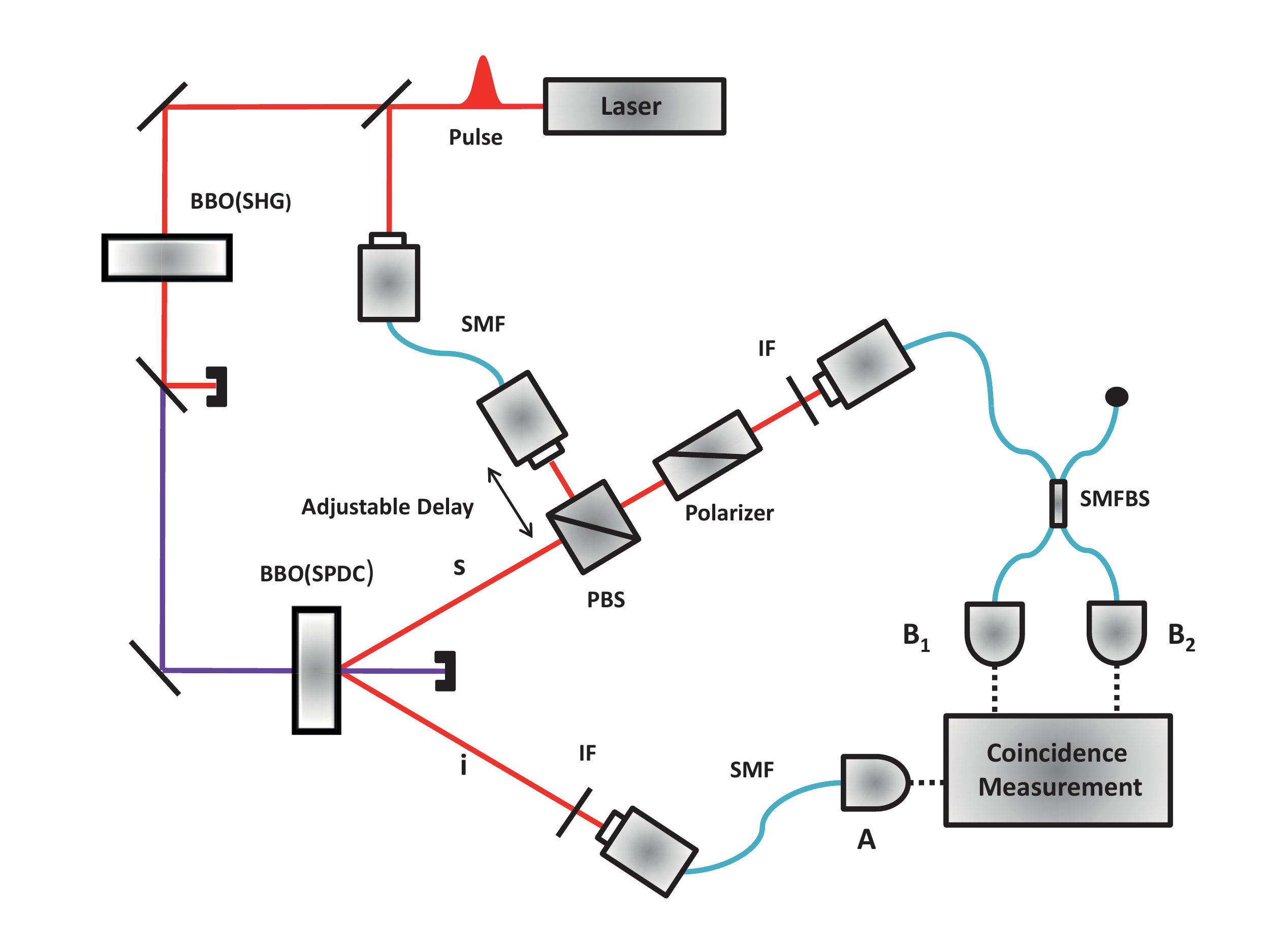}
\caption{Schematics for experimental setup to study the
indistinguishability induced photon statistical transition. SMF: single-mode
fiber; IF: interference filter; PBS: polarization beam splitter; SMFBS:
single-mode fiber beam splitter.}
\label{setup}
\end{figure}

In the experimental demonstration, we applied a single-photon state heralded from
spontaneous parametric down-conversion (SPDC). The coherent state was
directly from the attenuated laser. As shown in Fig.\ref{setup}, the $780$
\textrm{nm} pulsed laser beam was generated from a Ti:sapphire laser with a
repetition frequency of $76$ \textrm{MHz} and a pulse duration of $110$
\textrm{fs}. $390$ nm laser was obtained through a second harmonic generation
(SHG) process by a $\beta $-barium borate (BBO) crystal and served as the
pump light for type-$\mathrm{I}\mathrm{I}$ SPDC. The parametric light was
beam-like \cite{Takeuchi:01} and separated as the signal and idler beams.
The signal single-photon state can be heralded by detecting the idler
photon. Then, the single-photon state and the coherent state with orthogonal
polarizations were mixed together by a polarized beamsplitter. A
Glan--Thompson prism was used as a polarizer to project the orthogonal
polarized beams into a single polarization direction and removed the
distinguishable polarization information. Besides, the polarizer also
controlled the mixing ratio ($r$) with the rotation of polarization direction. $%
3$ \textrm{nm}-band-pass interference filters (IF) centered at $780$ \textrm{%
nm} and single mode fibers were used to ensure both spatial and temporal
modes overlapping for photons collection and interference.

\begin{figure}[t]
\centering
\includegraphics[width=7.5cm]{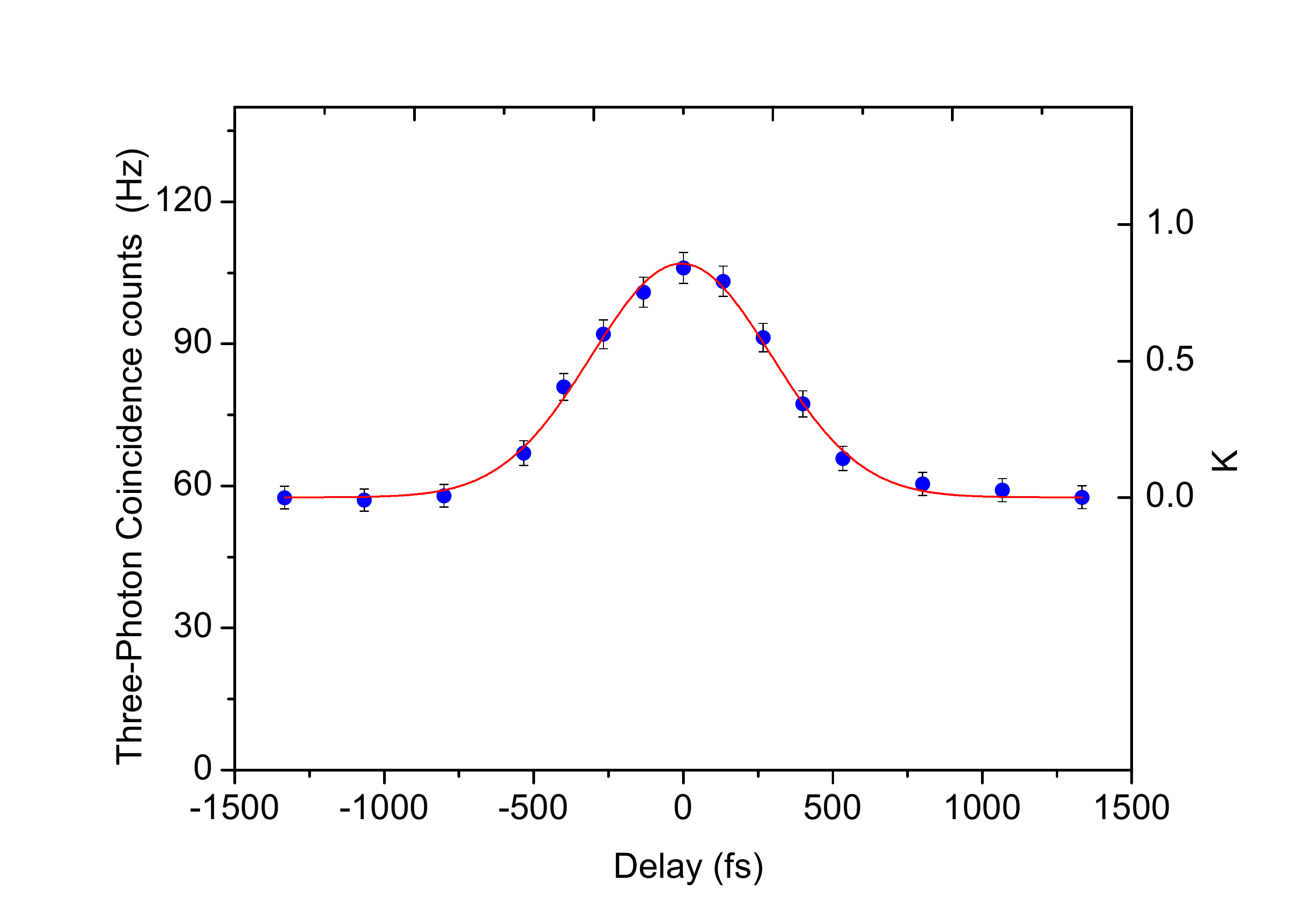}
\caption{Three-photon coincidence counts registered with different delays.
Dots are the experimental data. The error bars are given based on the total
coincidence counts. The solid line is the Gaussian fitting of the data.}
\label{De-Co}
\end{figure}

When indistinguishable photons from the coherent state and single-photon
state arrive at the PBS simultaneously, the photon bunching effect will
happen with more photon counts after the polarizer. Then,
the indistinguishability of the coherent state and single-photon state can be
measured from the enhancement of three-photon coincidence due to the
constructive interference \cite{HOM,Sun07}. By changing the relative delay
between the two mixed photon states, we can measure the three-photon
coincidence at zero delay ($N\left( 0\right) $) and the delay time much
longer than the pulse duration ($N\left( \infty \right) $), where the two
photon state are well separated. Therefore the value of
indistinguishability is
\begin{equation}
K=\frac{N\left( 0\right)}{N\left( \infty \right)} -1=0.86\pm 0.02\text{.}
\end{equation}%
The reason of $K<1$ may come from the imperfect overlapping of the spatial
and frequency modes and distinguishability of the single-photon state from
SPDC \cite{PhysRevA.79.013824}. Such a value can be further enhanced by
narrower interference filters. Also, by changing the delay time to control
the temporal overlapping between the two photon states, we were able to
modify $K$ in a simple way with $K\left( \tau \right) =N\left( \tau
\right) /N\left( \infty \right) -1$, where $K\left( \tau \right) $ and $%
N(\tau )$ represent the indistinguishability and the three-photon counts at
the delay of $\tau $. The experimental result is shown in Fig.\ref{De-Co}.
We can apply Gaussian distribution to fit the data as
\begin{equation}
K(\tau )=K\exp \left[ {-{{\left( {\frac{\tau }{\tau _{0}}}\right) }^{2}}}%
\right] \text{,}  \label{Kt}
\end{equation}%
where $\tau _{0}={425.1}\pm 11.6$ \textrm{fs}, which is determined by the duration of
the pump pulse, the band-width of the interference filter and the properties of
SPDC process in the BBO crystal, such as the thickness and phase matching
condition \cite{PhysRevA.79.013824,RaNC}. When $\tau \gg \tau _{0}$, the
two photon states was well temporally separated, $K(\tau )\rightarrow 0$.
Therefore we can modify the value of indistinguishability from $0$ to $K$
to study the photon indistinguishability and statistics.

\begin{figure}[tbph]
\centering
\includegraphics[width=7.5cm]{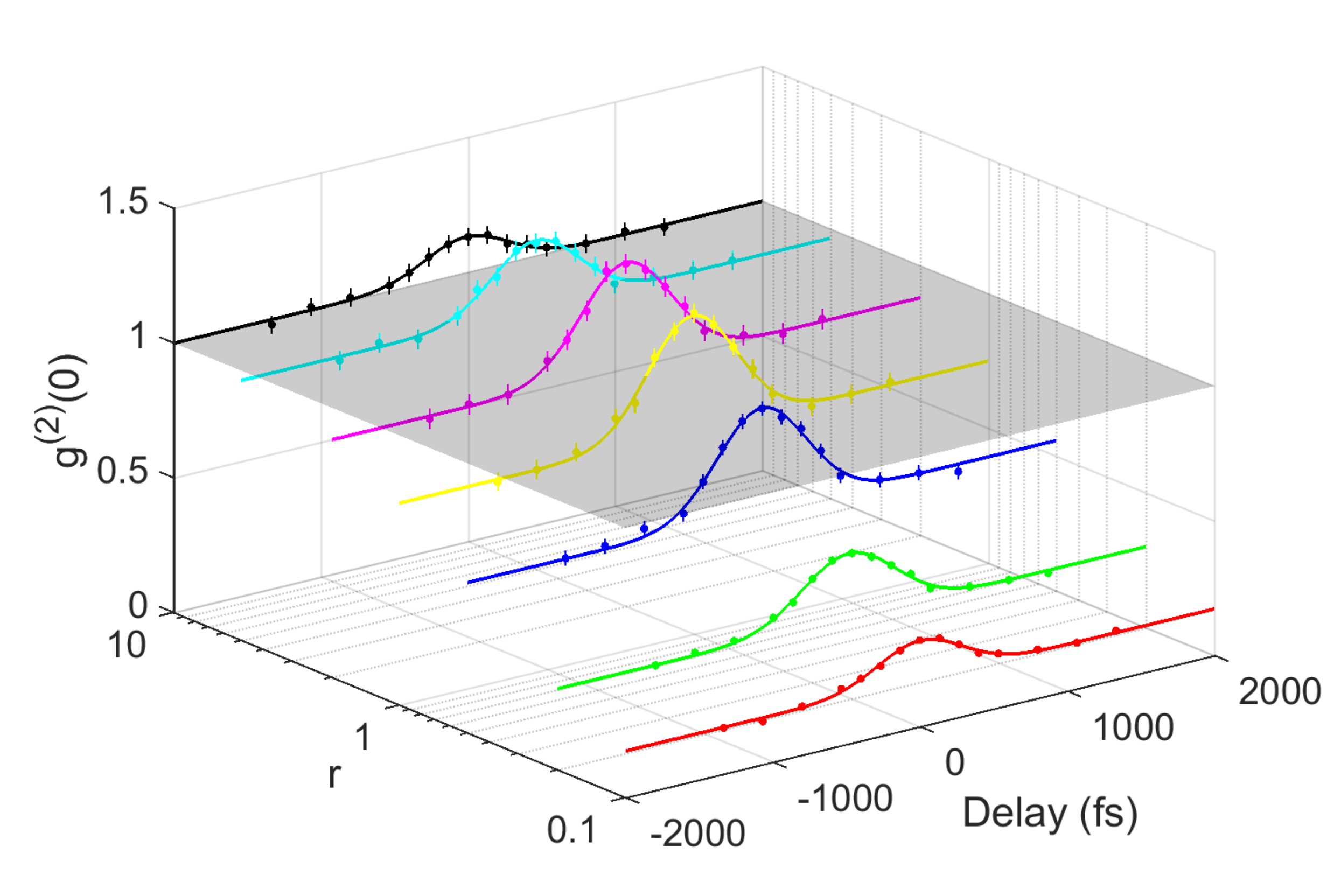}
\caption{The experimental values of ${g^{(2)}}\left( 0\right) $ of the mixed
photon state with mixing ratio $r$. Each colored curve is the Gaussian
fitting of the same colored dotted data.}
\label{3d}
\end{figure}

The ${g^{(2)}}\left( 0\right) $ of the mixed photon state can be measured
with the HBT interferometer, as shown in Fig.\ref{setup}. Since the
single-photon state was heralded by detecting the idler state, the ${g^{(2)}}%
\left( 0\right) $ of the mixed photon state at the signal path can be
obtained as:
\begin{equation}
{g^{(2)}}(0)=\frac{{{N_{A,{B_{1}},{B_{2}}}}}}{({{N_{A,{B_{1}}}}{N_{A,{B_{2}}%
}/{N_{A}^{2}})N_{A}}}}=\frac{{{N_{A,{B_{1}},{B_{2}}}}{N_{A}}}}{{{N_{A,{B_{1}}%
}}{N_{A,{B_{2}}}}}}\text{,}  \label{gtwo}
\end{equation}%
where ${N_{A}}$, ${N_{A,{B_{1}(B}_{2}{)}}}$, ${N_{A,{B_{1}},{B_{2}}}}$
represent the single-photon counts of detector $A$, two-photon coincidence
counts of detectors $A$ and ${B_{1}(B}_{2}{)}$, and three-photon coincidence
counts of detectors $A$, ${B_{1}}$ and ${B_{2}}$, respectively. By changing
photon indistinguishability ($K(\tau )$) with relative delay time and mixing
ratio ($r$) with the rotation of polarizer, we can modify the value of ${%
g^{(2)}}\left( 0\right) $ and the photon statistical behavior. Fig.\ref{3d} depicts
the results of a series measurement results. In each measurement, $r$ was
fixed, and the ${g^{(2)}}\left( 0\right) $ of the mixed photon state was
measured at different delays. The curves are the Gaussian fittings. Each of
them represents the value of ${g^{(2)}}\left( 0\right) $ as a function of $K$ with a certain mixed photon state. As the plane of ${g^{(2)}}\left( 0\right)
=1$, which is the boundary between classical and non-classical field, is
also presented in the figure, we can observe that some of the curves lie
across this plane. The experimental result manifests that the indistinguishability
is one of the main parameters of photon statistics and can lead to the
transition from classical (${g^{(2)}}\left( 0\right) >1$) to nonclassical (${%
g^{(2)}}\left( 0\right) <1$) regions.

\begin{figure}[t]
\centering
\includegraphics[width=7.5cm]{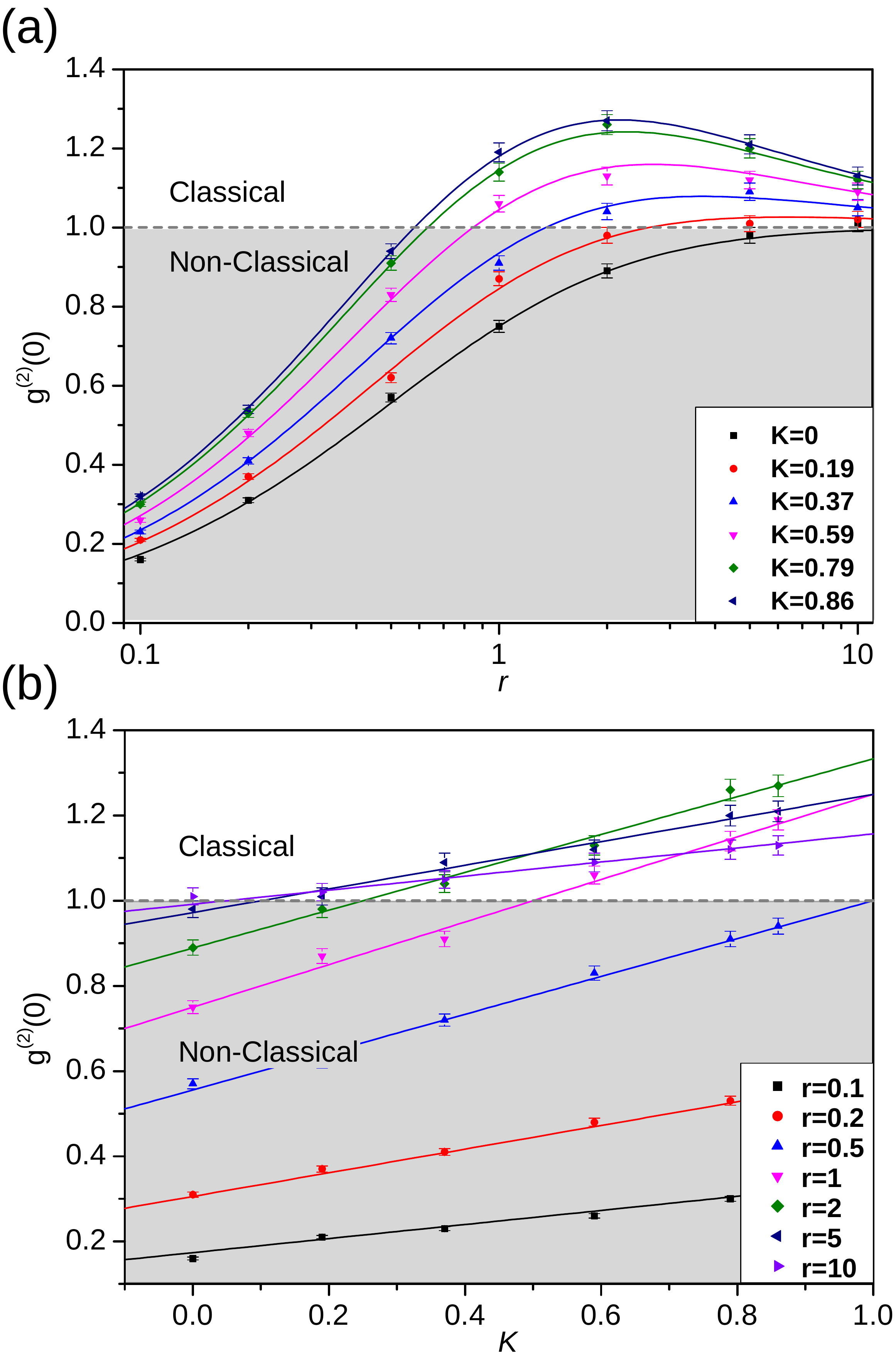}
\caption{The experimental value of ${g^{(2)}}\left( 0\right) $ with fixed
photon indistinguishability $K$ (a) and mixing ratio $r$ (b).
Experimental results are shown in color dots with error bars and solid color
lines are fittings with Eq.(\protect\ref{gtwo3}).}
\label{g2}
\end{figure}

By converting time delay ($\tau$) to photon indistinguishability ($K$) with Eq.(\ref%
{Kt}), we can demonstrate the photon statistics behavior with different $K$
and $r$. Fig.\ref{g2}(a)(b) shows the value of ${g^{(2)}}\left( 0\right) $
with fixed photon indistinguishability $K$ and mixing ratio $r$,
respectively. Each sequence of colored dots represents the measured value
and the colored line is the theoretical result from Eq.(\ref{gtwo3}). The
nonclassical photon statistics with ${g^{(2)}}\left( 0\right) <1$ is shown
in the grey area. In Fig.\ref{g2}(a), when $r$ is much smaller than $1$,
the single-photon state from SPDC dominates the photon statistics, demonstrating
the non-classical behavior. When $r$ is much larger than $1$, the coherent
state dominates the photon statistics with ${g^{(2)}}\left( 0\right) =1$.
For these two states, the values of ${g^{(2)}}\left( 0\right) $ are never
larger than $1$. However, when they are mixed with some ratios, the photon
indistinguishability induced photon bunching will significantly change the photon
statistics, demonstrating the transition to a classical behavior with ${g^{(2)}}\left(
0\right) >1$. Fig.\ref{g2}(b) clearly shows the contribution of the photon
indistinguishability to the photon statistics. When the photons are totally
distinguishable ($K=0$), the ${g^{(2)}}\left( 0\right) $ of the mixed state
is always less than $1$. However, when $K$ increases, the value of ${g^{(2)}}%
\left( 0\right) $ increases to be larger than $1$ and a transition from the
nonclassical to classical field happens when $r>0.5$.

In conclusion, we have studied the photon indistinguishability induced
photon bunching effect to modify the photon statistics. In a photon
interference, the photon indistinguishability is changed by the temporal
overlapping of the single photon state and the weak coherent state. The
transition from a classical to nonclassical photon statistical behavior was
experimentally demonstrated by changing the indistinguishability of photons.
It provides a new method to manipulate the photon statistics for the study
of quantum coherence. Besides the quantum phase, the study also indicates
that the photon indistinguishability is a key parameter for quantum
coherence, especially for multi-partite quantum coherence.

\section*{Acknowledgment}

This work was supported by the Strategic Priority Research Program(B) of the
Chinese Academy of Sciences (Grant No. XDB01030200), the National Natural
Science Foundation of China (Nos. 61522508, 11374290, 91536219, 11504363),
the Fundamental Research Funds for the Central Universities.

\appendix
\section{The calculation of ${g^{2}}\left( 0\right)$}
In this appendix, we show the calculation of ${g^{2}}\left( 0\right) $. The
coherent state is described as
\begin{equation}
\left\vert \alpha \right\rangle =\exp \left( {-{{{{\left\vert \alpha
\right\vert }^{2}/}}2}}\right) \sum\limits_{n^{\prime }=0}^{\infty }{{\frac{{%
{\alpha ^{n^{\prime }}}}}{{{{\left( {n}^{\prime }{!}\right) }^{{1/2}}}}}}}%
\left\vert {n}^{\prime }\right\rangle \text{.}
\end{equation}%
To distinguish the photons from single-photon state and coherent state, here
the number state from coherent state is written as $\left\vert {n^{\prime }}%
\right\rangle $. Therefore, the single-photon added coherent state can be
described as in Eq.(\ref{rho3}) with $\mathrm{e}^{{-{{\left\vert \alpha
\right\vert }^{2}}}}\approx 1/(1+{{{\left\vert \alpha \right\vert }^{2})}}%
\approx 1$ when ${\left\vert \alpha \right\vert ^{2}}\ll 1$. The amplitude
${1+K}$ is proportional to the two-photon counts of $\left\vert {1,1^{\prime
}}\right\rangle \left\langle {1,1^{\prime }}\right\vert $  \cite%
{PhysRevA.79.013824}:
\begin{eqnarray}
C_{2} &=&\left\langle {1,1^{\prime }}\right\vert a^{\dag }a^{\dag
}aa\left\vert {1,1^{\prime }}\right\rangle  \notag \\
&=&2+2\mathrm{Tr}(\left\vert 1\right\rangle \left\langle 1\right\vert
\left\vert 1^{\prime }\right\rangle \left\langle 1^{\prime }\right\vert ) \notag \\
&=&2(1+K)\text{.}
\label{C2}
\end{eqnarray}
If $K=1$ with $\left\vert 1\right\rangle \equiv \left\vert 1^{\prime
}\right\rangle $, the indistinguishable two-photon state is $2\left\vert
2\right\rangle {\left\langle 2\right\vert }$, showing an
indistinguishability induced perfect photon bunching effect with $(a^{\dag
})^{2}{\left\vert 0\right\rangle =}\sqrt{2}\left\vert 2\right\rangle $.
However, if $\left\vert 1\right\rangle \perp \left\vert 1^{\prime
}\right\rangle $, $K=0$, $a^{\dag }a^{\prime \dag }{\left\vert
0\right\rangle =}\left\vert 1\right\rangle \otimes \left\vert 1^{\prime
}\right\rangle $, the two-photon state is $\left\vert 1\right\rangle
\left\langle 1\right\vert \otimes \left\vert 1^{\prime }\right\rangle
\left\langle 1^{\prime }\right\vert $. For partially distinguishable cases
with $0<K<1$, the enhancement of $K$ in two-photon counts describes an
imperfect two-photon bunching effect \cite{PhysRevA.79.013824}.

When ${\left\vert \alpha \right\vert ^{2}}\ll 1$, we can omit high order
terms in the calculation of the second correlation function. Based on Eq.(\ref{C2}), we can get
\begin{eqnarray}
g^{\left( 2\right) }\left( 0\right)  &=&{\frac{{\left\langle {{a^{\dag }}{%
a^{\dag }}aa}\right\rangle }}{{{{\left\langle {{a^{\dag }}a}\right\rangle }%
^{2}}}}}  \notag \\
&=&{\frac{\mathrm{Tr}{\left( {{a^{\dag }}{a^{\dag }}aa{\rho _{K}}}\right) }}{%
\mathrm{Tr}{{{\left( {{a^{\dag }}a{\rho _{K}}}\right) }^{2}}}}}  \notag \\
&=&{\frac{{\left( {1-\eta }\right) {\alpha ^{4}}+2\eta \left( {1+K}\right) {%
\alpha ^{2}}}}{{{{\left[ {\left( {1-\eta }\right) \alpha +\eta }\right] }^{2}%
}}}}  \notag \\
&=&{\frac{{\left( {1-\eta }\right) }r{{^{2}}+2\left( {1+K}\right) {r}}}{[{{{{%
\left( {1-\eta }\right) r+1]}}^{2}}}}}\text{,}  \label{gk}
\end{eqnarray}%
where $r\equiv {{\alpha ^{2}/}\eta }$.

Under the assumption that $\eta \ll 1$, $1-\eta \approx 1$, Eq.(\ref{gk})
can be simplified as Eq.(\ref{gtwo3})
\begin{equation}
{g^{2}}\left( 0\right) ={\frac{{{r^{2}}+2\left( {1+K}\right) r}}{{{{\left( {%
1+r}\right) }^{2}}}}}\text{.}
\end{equation}

\end{document}